# Upper and lower limits for the number of S-wave bound states in an attractive potential

Fabian Brau[a]

*Service de Physique Générale et de Physique des Particules Elémentaires, Groupe de Physique Nucléaire Théorique, Université de Mons-Hainaut, Mons, Belgique*

Francesco Calogero[b]

*Dipartimento di Fisica, Università di Roma "La Sapienza" and Istituto Nazionale di Fisica Nucleare, Sezione di Roma, Rome, Italy*

*Summary*

New upper and lower limits are given for the number of S-wave bound states yielded by an attractive (monotonic) potential in the context of the Schrödinger or Klein-Gordon equation.

[a] FNRS Postdoctoral Researcher. E-mail : fabian.brau@umh.ac.be
[b] E-mail : francesco.calogero@roma1.infn.it, francesco.calogero@uniroma1.it

# I. INTRODUCTION AND MAIN RESULTS

The determination in the framework of nonrelativistic quantum mechanics of necessary and sufficient conditions for the existence of bound states in a given potential and, more generally, of upper and lower limits for the number of bound states yielded by such a potential, has engaged the attention of theoretical and mathematical physicists since the early fifties, and, notwithstanding the fact that, with modern computers, the numerical evaluation of the number of bound states for a given potential is an easy task, it continues to be actively pursued: see for instance [1-23], as well as the surveys of (some of) these results in [24-25]. In this paper we provide new upper and lower limits for the number of S-wave bound states possessed by a central potential vanishing at infinity and yielding a nowhere repulsive force and we compare them, for some test potentials, with the exact results and with previously known upper and lower limits. These comparisons indicate that these new limits are generally more stringent than hitherto known results and indeed remarkably cogent, especially for potentials possessing many bound states.

Let us briefly review (some of) the previous findings, focusing on those relevant to our treatment, hence restricting attention to the S-wave case (even when results are also known for higher partial waves). Hereafter – except in Section IV – we use the standard nonrelativistic quantum mechanical units such that $\hbar^2/(2m) = 1$, which entail that the potential $V(r)$ has the dimension of an inverse square length, and we indicate with $N$ the number of S-wave bound states. We also assume throughout that the potential $V(r)$ is less singular than the inverse square radius at the origin and that it vanishes asymptotically faster than the inverse square radius, say (for some *positive* $\varepsilon$)

$$\lim_{r \to 0} \left[ r^{2-\varepsilon} V(r) \right] = 0, \tag{1.1a}$$

$$\lim_{r \to \infty} \left[ r^{2+\varepsilon} V(r) \right] = 0. \tag{1.1b}$$

Note that these assumptions entail that the square root of the (modulus of the) potential is integrable both at the origin and at infinity.

Bargmann [2] and Schwinger [3] obtained the following upper limit for $N$:

$$\text{BS:} \quad N \leq \int_0^\infty dr \, r |V(r)|. \tag{1.2}$$

This result is generally referred to as the Bargmann-Schwinger bound; we hereafter refer to it as the BS (upper) limit. This result was obtained after Jost and Pais [1] had shown that the fact that the right-hand side of (1.2) exceed unity is a *necessary* condition for the existence of bound states (namely, the special case of the BS limit with $N = 1$).

Cohn [5,6] and Calogero [7,8] later obtained another upper limit for $N$, which is valid provided the force associated with the potential $V(r)$ is nowhere repulsive, namely the potential $V(r)$ is a monotonically nondecreasing function of the radius $r$,

$$dV(r)/dr \geq 0, \tag{1.3}$$

entailing of course that the potential is everywhere negative, $V(r) = -|V(r)|$. This upper limit reads:

$$\text{CC:} \quad N \leq \frac{2}{\pi} \int_0^\infty dr \, |V(r)|^{1/2}. \tag{1.4}$$

This result has been referred to as the Calogero-Cohn bound [21]; hereafter we shall refer to it as the CC (upper) limit. This CC limit, (1.4), in contrast to the BS limit (1.2), features the correct dependence on the strength of the potential; indeed it has been shown [24] that, for any potential $V(r)$, if a measure of the strength of the potential is introduced via the introduction of a "coupling constant" $g^2$ by setting

$$V(r) = g^2 v(r), \tag{1.5}$$

then as $g$ diverges to positive infinity, $N$ grows proportionally to $g$. But it is also known [12] (see also [8,11,26]) that asymptotically, as $g$ diverges, $g \to \infty$,

$$N \approx \frac{1}{\pi}\int_0^\infty dr\,|V(r)|^{1/2} = \frac{g}{\pi}\int_0^\infty dr\,|v(r)|^{1/2}\,. \tag{1.6}$$

Hence for strongly attractive potentials featuring many bound states the CC limit (1.4) tends to overestimate $N$ by a factor $2$. The main merit of the new limits provided in this paper is to remedy this defect (see below).

Some modifications of the inequality (1.4) and of the condition (1.3) on the shape of the potential have been introduced by Chadan *et al.* [21]. These modifications lead to less restrictive inequalities but more flexible conditions on the shape of the potential, allowing for some oscillations.

Another upper bound, which also gives the correct power behavior of the number of bound states when the strength of the potential diverges, has been obtained by Martin [16]:

$$\text{M:}\quad N \leq \left[\int_0^\infty dr\,r^2 V^-(r) \int_0^\infty dr\,V^-(r)\right]^{1/4}, \tag{1.7}$$

where $V^-(r)$ is the *negative* part of $V(r)$. This limit is applicable even if the potential does not satisfy the property to yield a nowhere repulsive force, see (1.3), but it is nontrivial only for potentials the nonpositive part of which is integrable at the origin. Hereafter we refer to it as the M (upper) limit.

The known lower limits on $N$ are scarcer and less neat. A result [8] states that

$$N \geq \frac{1}{\pi}\int_0^\infty dr\,\min[1/a,\,-aV(r)] - \frac{1}{2}, \tag{1.8a}$$

where $a$ is an arbitrary *positive* constant, $a > 0$, and

$$\min[x,y] = x\quad \text{if}\quad x \leq y,\quad \min[x,y] = y\quad \text{if}\quad y \leq x\,. \tag{1.8b}$$

By choosing $a$ proportional to $g^{-1}$ (see (1.5)) it is clear that this limit has the correct power growth when $g$ diverges. The most stringent version of this limit obtains by performing firstly the integration in the right-hand side of (1.8a), and by then maximizing the result over all *positive* values of the parameter $a$. For everywhere nondecreasing potentials, see (1.3), the minimum definition (1.8b) is easily implemented by splitting the integration range in (1.8a) in two parts, and thereby, via standard computations, one arrives at the somewhat neater lower limit

$$\text{C:}\quad N \geq \frac{2}{\pi}\rho|V(\rho)|^{1/2} - \frac{1}{2}, \tag{1.9a}$$

where $\rho$ is a root of the equation

$$\rho V(\rho) = \int_\rho^\infty dr\,V(r)\,. \tag{1.9b}$$

This limit will be hereafter referred to as the C (lower) limit.

If the potential, besides satisfying the monotonicity condition (1.3), is finite at the origin, a more explicit if less cogent result obtains by setting $a = |V(0)|^{-1/2}$ in (1.8):

$$\text{C}_0\text{:}\quad N \geq \frac{1}{\pi}\int_0^\infty dr\,\frac{|V(r)|}{|V(0)|^{1/2}} - \frac{1}{2}\,. \tag{1.10}$$

Hereafter we shall refer to this result as the $C_0$ (lower) limit.

By setting $N = 1$ in (1.8), (1.9) respectively (1.10) one obtains the following three conditions each of which is *sufficient* to guarantee the existence of (at least) one bound state:

$$\int_0^\infty dr\,\min[1/a,\,-aV(r)] > \frac{3\pi}{2}, \tag{1.11a}$$

$$\rho |V(\rho)|^{1/2} > \frac{3\pi}{4} \qquad (1.11b)$$

with $\rho$ again a root of (1.9b),

$$\int_0^\infty dr\, |V(r)| > \frac{3\pi}{2} |V(0)|^{1/2} . \qquad (1.11c)$$

In the first of these inequalities, (1.11a), $a$ is an arbitrary *positive* constant; the most stringent condition obtains of course by performing firstly the integration in the left-hand side and by then minimizing the result over all *positive* values of $a$; the other two inequalities, (1.11b,c), are neater but for their validity it is required that the potential satisfies the monotonicity condition (1.3) (and of course (1.11c) is only applicable if the potential is finite at the origin).

In view of future applications (see below) let us also report two other conditions which are *sufficient* to guarantee that the potential $V(r)$ possesses (at least) one bound state [6, 24]:

$$a^{-1} \int_0^a dr\, r^2 |V(r)| + a \int_a^\infty dr\, |V(r)| > 1 , \qquad (1.12)$$

$$a \int_0^\infty dr\, |V(r)| / \left[1 + a^2 |V(r)|\right] > 1 . \qquad (1.13)$$

Both these conditions apply provided the potential is nowhere positive, $V(r) = -|V(r)|$; in both of them $a$ is an arbitrary *positive* constant, and of course the most stringent conditions obtain by minimizing the left-hand sides over all *positive* values of $a$. It is easily seen that, in the case of (1.12), the minimizing value of $a$ is the root of the equation

$$\int_0^a dr\, r^2 |V(r)| = a^2 \int_a^\infty dr\, |V(r)| , \qquad (1.14)$$

(entailing that the two terms in the left-hand side of (1.12) yield equal contributions), in the case of (1.13) it is the root of the equation

$$\int_0^\infty dr\, |V(r)| \left(1 - a^2 |V(r)|\right) \left(1 + a^2 |V(r)|\right)^{-2} = 0 . \qquad (1.15)$$

After this terse survey of previous results let us now report the new upper and lower limits on the number $N$ of S-wave bound states obtained in this paper, in which we restrict for simplicity attention to potentials that satisfy the monotonicity condition (1.3) (we plan to report results applicable to more general potentials, as well as to higher partial waves, in a subsequent paper). These limits are of two different types.

The (new) upper limit of the first type reads as follows:

$$N \leq \frac{1}{\pi} \int_0^\infty dr\, |V(r)|^{1/2} + \frac{1}{4\pi} \log\left|\frac{V(p)}{V(q)}\right| + \frac{1}{2} , \qquad (1.16a)$$

with the two distances $p$ and $q$ defined by the relations

$$\int_0^p dr\, |V(r)|^{1/2} = \pi/2 , \qquad (1.16b)$$

$$\int_q^\infty dr\, |V(r)|^{1/2} = \pi/2 . \qquad (1.16c)$$

Clearly these two formulas, (1.16b) respectively (1.16c), provide an unambiguous definition of the two quantities $p$ respectively $q$, provided the potential $V(r)$ possesses at least one bound state, since it must then satisfy the following *necessary* condition for the existence of bound states [7] (corresponding to (1.4) with $N = 1$):

$$\int_0^\infty dr \, |V(r)|^{1/2} \geq \pi/2 \, . \tag{1.17}$$

And also note that, due to the assumed monotonicity of the potential, see (1.3), (1.16) entails that a neater albeit less stringent upper limit to $N$ is provided by the formula

$$N \leq \frac{1}{\pi} \int_0^\infty dr \, |V(r)|^{1/2} + \frac{1}{4\pi} \log \left| \frac{V(0)}{V(q)} \right| + \frac{1}{2} \, , \tag{1.18}$$

with $q$ always defined by (1.16c). This upper limit is however nontrivial only for potentials that are finite at the origin.

The (new) lower limit of the first type reads (for potentials that are finite at the origin)

$$N > \frac{1}{\pi} \int_0^s dr \, |V(r)|^{1/2} - \frac{1}{4\pi} \log \left| \frac{V(0)}{V(s)} \right| - \frac{1}{2} \, , \tag{1.19}$$

with $s$ an arbitrary (of course *positive*) radius. The choice of $s$ that produces the most stringent bound is the root of the following nondifferential equation in $s$ (here, and always below, appended primes denote differentiations):

$$V'(s) = 4 |V(s)|^{3/2} \, . \tag{1.20}$$

Indeed, the values of $s$ which satisfy this last equation maximize the right hand side of (1.19). If this equation possesses more than one positive root, generally the most stringent bound obtains by choosing the largest.

A neater, if generally less stringent, lower bound obtains by choosing $s = q$, since via (1.16c) one then gets

$$N > \frac{1}{\pi} \int_0^\infty dr \, |V(r)|^{1/2} - \frac{1}{4\pi} \log \left| \frac{V(0)}{V(q)} \right| - 1 \, , \tag{1.21}$$

The analogy of this formula, (1.21), to (1.16) is remarkable; and of course this lower limit to $N$ is also nontrivial only if the potential $V(r)$ is finite at the origin.

If the potential is singular at the origin a neat lower bound, analogous to (1.16a), reads

$$N \geq \frac{1}{\pi} \int_0^\infty dr \, |V(r)|^{1/2} - \frac{1}{4\pi} \log \left| \frac{V(p)}{V(q)} \right| - \frac{3}{2} \, , \tag{1.22}$$

with $p$ and $q$ defined by (1.16b,c).

A less neat but generally more stringent (albeit only marginally so) lower bound that looks somewhat analogous to (1.19) and is also applicable to potentials that are singular at the origin reads

$$N > \frac{1}{\pi} \int_t^s dr \, |V(r)|^{1/2} - \frac{1}{4\pi} \log \left| \frac{V(p)}{V(s)} \right| \, , \tag{1.23a}$$

with $p$ defined by (1.16b) and $s \geq t$ but otherwise *arbitrary*. As for the *positive* quantity $t$, a characterization of it adequate to guarantee validity of this lower limit, (1.23a), is the requirement that it be the smallest positive root of the (nondifferential) equation

$$t = \int_0^t dr \, r^2 |V(r)| \, . \tag{1.23b}$$

Another characterization of $t$, which leads to a (generally only marginally) more stringent lower limit, is provided in Section III. Note that, as above, the choice of $s$ in (1.23a) that yields the most stringent bound is the root of the nondifferential equation (1.20) (provided of course such a choice of $s$ is compatible with the condition $s \geq t$, as it is certainly the case for strong potentials possessing many bound states). And again, as above, a neater, if generally less stringent, lower bound obtains by choosing $s = q$, since via (1.16c) one then gets, in place of (1.23a),

$$N > \frac{1}{\pi}\int_t^\infty dr |V(r)|^{1/2} - \frac{1}{4\pi}\log\left|\frac{V(p)}{V(q)}\right| - \frac{1}{2}, \qquad (1.23c)$$

again of course with $q$ respectively $t$ defined by (1.16b) and (1.23b) (of course provided $q \geq t$, as it is certainly the case for strong potentials).

Let us now report a second type of (new) limits on the number $N$ of S-wave bound states, which are particularly suitable for numerical computations, although there exist also cases amenable to analytic treatment (see Section II).

Firstly we report an upper limit, valid for potentials finite at the origin, to which consideration is, for simplicity, here restricted. Let us define the radius $q$ via (1.16c), and the sequence of *increasing* radii $r_j^{(+)}$ via the explicit recursion relation

$$r_{j+1}^{(+)} = r_j^{(+)} + (\pi/2)\left|V\left(r_j^{(+)}\right)\right|^{-1/2}, \quad r_0^{(+)} = 0; \qquad (1.24)$$

and let the positive integer $J^{(+)}$ be defined by the condition that the radius $r_{J^{(+)}+1}^{(+)}$ yielded by this recursion (be the first one to) exceed or equal $q$,

$$r_{J^{(+)}}^{(+)} < q \leq r_{J^{(+)}+1}^{(+)}. \qquad (1.25)$$

The upper limit is then provided by the inequality

$$N \leq \{\!\{(J^{(+)}+1)/2\}\!\} + 1. \qquad (1.26)$$

Here and always below the double braces denote the integer part, $\{\!\{J/2\}\!\} = J/2$ if $J$ is *even*, $\{\!\{J/2\}\!\} = (J-1)/2$ if $J$ is *odd*.

Finally we report an analogous lower limit to $N$, which does not require that $V(r)$ be finite at the origin to yield a nontrivial result. Again, one first defines the radius $q$ via (1.16c), and then introduces a series of *decreasing* radii $r_j^{(-)}$ via the explicit recursion relation

$$r_{j+1}^{(-)} = r_j^{(-)} - (\pi/2)\left|V\left(r_j^{(-)}\right)\right|^{-1/2}, \quad r_0^{(-)} = q. \qquad (1.27)$$

Now let the positive integer $J^{(-)}$ be defined by the condition that the quantity $r_{J^{(-)}}^{(-)}$ yielded by this recursion be the last one to be *positive*,

$$r_{J^{(-)}+1}^{(-)} \leq 0 < r_{J^{(-)}}^{(-)}. \qquad (1.28)$$

The lower limit is then provided by the inequality

$$N \geq \{\!\{J^{(-)}/2\}\!\}. \qquad (1.29)$$

In Section II we provide several tests of the efficacy of our upper and lower limits; in Section III, we prove them; in Section IV we point out that all the results reported herein in the (nonrelativistic) context of the Schrödinger equation can be easily extended to the (kinematically relativistic, if only first-quantized) Klein-Gordon case.

## II.   TESTS

Most of the limits on the number of S-wave bound states reported in the preceding Section I are "best possible", namely it is generally possible to find potentials that saturate them. The shape of these saturating potentials can generally be easily inferred from the very procedure whereby the limits were derived; in particular for our new limits the saturating potentials are generally of ladder type (including the simplest such potential, the square-well), since for such potentials the second term in the right-hand side of (3.7) tends to vanish (as discussed in some detail in the last part of the following Section III). But while the fact that the formula providing a limit has the property to be "best possible" entails that there can be no hope to make it more stringent by just modifying some constant appearing in it (it is for instance impossible to obtain a more stringent upper limit than (1.4) by just replacing the constant $2/\pi$ in the right-hand side by a smaller number), it does by no means imply that such a bound provides a stringent limitation for all potentials; far from it (as we will presently see). Indeed, a more interesting question is how different limits behave for a variety of (test) potentials. This Section is devoted to such an assessment, for which we use six different potentials: the square-well potential (hereafter referred to as SW)

$$\text{SW: } V(r) = -g^2 R^{-2} \quad \text{for} \quad r \leq R, \tag{2.1a}$$

$$\text{SW: } V(r) = 0 \quad \text{for} \quad r > R; \tag{2.1b}$$

the Pöschl-Teller [27] (or "single-soliton", see for instance [28]) potential (hereafter referred to as PT),

$$\text{PT: } V(r) = -g^2 R^{-2} \left[\cosh(r/R)\right]^{-2}; \tag{2.2}$$

the exponential potential (hereafter referred to as E),

$$\text{E: } V(r) = -g^2 R^{-2} \exp(-r/R); \tag{2.3}$$

the Hulthén potential (hereafter referred to as H),

$$\text{H: } V(r) = -g^2 R^{-2} \left[\exp(r/R) - 1\right]^{-1}; \tag{2.4}$$

the Yukawa potential (hereafter referred to as Y),

$$\text{Y: } V(r) = -g^2 (rR)^{-1} \exp(-r/R); \tag{2.5}$$

and the following Shifted and Truncated Inverse Square potential (hereafter referred to as STIS), which has the merit to allow analytic computation of all limits as well as of the exact number of bound states (see below):

$$\text{STIS: } V(r) = -g^2 (R+r)^{-2} \quad \text{for} \quad 0 \leq r \leq \alpha R, \tag{2.6a}$$

$$\text{STIS: } V(r) = 0 \quad \text{for} \quad r > \alpha R. \tag{2.6b}$$

In all these equations, and below, $R$ is an arbitrary (of course *positive*) given radius, and $g$, as well as $\alpha$ in the last equation, (2.6), are arbitrary dimensionless *positive* constants.

We only report, for the new limits of the first type, tests of the *neatest* limits given in the previous Section I, namely we consider the upper respectively lower limits (1.18) respectively (1.21) for regular potentials, and the upper respectively lower limits (1.16) respectively (1.22) (only) for singular potentials; indeed, for regular potentials, the difference between the neater upper limit (1.18) and the more stringent upper limit (1.16) is generally negligibly small (namely, less than one unit), and likewise for the difference between the neater lower limit (1.21) and the more stringent lower limits (1.19) or (1.23). (Let us however emphasize that when one considers potentials with few bound states or searches for constraints on potential parameters necessary or sufficient for the existence of one bound state, it is advisable to use the most stringent available limits). As for the new limits of the second type, we test the upper respectively lower limits (1.26) respectively (1.29) for regular potentials, and the lower limit (1.29) for singular potentials. The tests are performed by comparing the new limits with the exact results, and with the previously known limits reported (and named) in Section I.

The simplest test is provided by the (nonsingular) SW potential (2.1), for which the exact number of bound states is given by the formula

$$N = \{\!\{\nu\}\!\}, \tag{2.7}$$

with

$$\nu = \frac{g}{\pi} + \frac{1}{2}. \tag{2.8}$$

In this case the new limits obtained in this paper tend to give the exact result (as explained above), except for the approximations introduced in order to obtain neater formulas. Indeed the upper respectively lower limits of the first type (1.18) respectively (1.21) (with $q < R$ as implied by (1.16c), so that the logarithmic terms in both these formulas vanish) yield $N \leq \nu$ respectively $N > \nu - 3/2$, while the more stringent lower bound (1.19) with $s = R$ yields $N \geq \nu - 1$. The upper respectively lower limits of the second type, (1.26) respectively (1.29), can as well be computed analytically for this potential, yielding $N \leq \nu + 1/2$ respectively $N \geq \nu - 1$. The BS, CC and M upper limits do not produce such good results. The BS upper limit yields $N \leq g^2/2$, which gives a very poor limitation when $g$ (hence the number of bound states) grows (indeed we know that the BS upper limit is always very poor for strong potentials, see also below). The CC respectively M upper limits do give the correct linear behavior in $g$, but with too big a slope, respectively $N \leq 2g/\pi = 2(\nu - 1/2)$ and $N \leq 3^{-1/4} g = 3^{-1/4} \pi (\nu - 1/2) = 2.387(\nu - 1/2)$. Finally, in this particular case the C and $C_0$ lower limits coincide and yield $N \geq \nu - 1$, namely a slightly more stringent limit than (1.21) (indeed, just the same result as (1.19), see above).

The second test is performed with the (nonsingular) PT potential (2.2). For this potential the exact number of bound states is again given by (2.7) but now with

$$\nu = \left(\sqrt{1 + 4g^2} + 1\right)/4, \qquad (2.9a)$$

which, in the limit of large $g$, yields

$$\nu = \frac{1}{2}g + \frac{1}{4} + \frac{1}{16g} + O(g^{-3}). \qquad (2.9b)$$

In this case the new upper and lower limits of the first type, (1.18) respectively (1.21), can as well be computed analytically, and they read

$$N \leq \frac{g}{2} - \frac{1}{2\pi} \log\left[\sin\left(\frac{\pi}{2g}\right)\right] + \frac{1}{2}, \qquad (2.10a)$$

respectively

$$N > \frac{g}{2} + \frac{1}{2\pi} \log\left[\sin\left(\frac{\pi}{2g}\right)\right] - 1, \qquad (2.11a)$$

entailing, in the limit of large $g$,

$$N \leq \frac{g}{2} + \frac{1}{2\pi} \log\left(\frac{2g}{\pi}\right) + \frac{1}{2} + \frac{1}{12\pi}\left(\frac{\pi}{2g}\right)^2 + O(g^{-4}), \qquad (2.10b)$$

respectively

$$N > \frac{g}{2} - \frac{1}{2\pi} \log\left(\frac{2g}{\pi}\right) - 1 - \frac{1}{12\pi}\left(\frac{\pi}{2g}\right)^2 + O(g^{-4}). \qquad (2.11b)$$

As for the new limits of the second type, (1.26) and (1.29), in this case they can only be evaluated numerically. In Fig. 1 we present, for this potential, a comparison between the exact number of bound states, the new limits of the first and of the second type, and the previously known C, $C_0$ lower limits, and BS, CC, M upper limits, all of which can be computed analytically: BS: $N \leq \log(2) g^2$ (very bad at large $g$); CC: $N \leq g$; M: $N \leq (\pi^2/12)^{1/4} g \approx 0.95 g$ (both of which give roughly twice the correct result at large $g$); C: $N \geq (2/\pi) \exp(-x) g - 1/2 \approx 0.336 g - 1/2$ (where $x$ is the root of $2x = 1 + \exp(-2x)$); $C_0$: $N \geq g/\pi - 1/2 \approx 0.318 g - 1/2$ (the C and the $C_0$ lower bounds are less stringent than the lower bound (2.11) as soon as $g$ exceeds 3.98 and 3.48 respectively). As it is clear from Fig. 1, the new bounds are quite cogent; and from (2.10b) and (2.11b) one sees that those of the first type remain quite stringent as well for rather large values of $g$: for instance when the exact number $N$ of bound states is equal to 5000 these upper and lower limits restrict it to the rather small interval $[4998, 5001]$. Likewise, at this value of $g$, the new limits of the second type, (1.26) respectively (1.29), entail the restrictions $4996 \leq N \leq 5002$; while the corresponding value of the BS upper limit exceeds $6.9 \times 10^7$, the CC upper limit only informs us that $N \leq 10^4$, and the lower limit C that $N \geq 3360$.

**Figure 1**. *Comparison between the exact number of bound states for the PT potential (2.2) (ladder curve), the bounds of the first type (1.18) and (1.21), the C and $C_0$ lower bounds, the BS, CC and M upper bounds, and the bounds of the second type (1.26) and (1.29).*

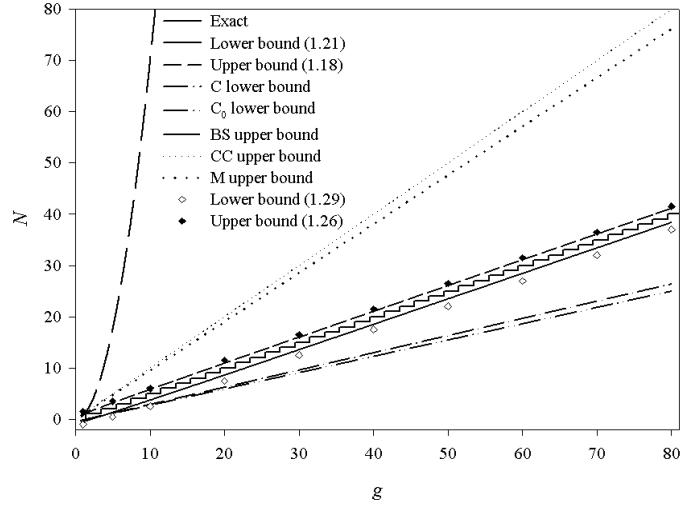

The third test is performed with the (regular) E potential (2.3). In this case the exact number of bound states coincides with the number of zeros of the zeroth-order Bessel function $J_0(x)$ in the interval $0 < x \leq 2g$ (see for example [29, p. 196]). Also in this case the new upper and lower limits of the first type (1.18) and (1.21) can be computed analytically:

$$N \leq \frac{2}{\pi}g + \frac{1}{2\pi}\log\left(\frac{4}{\pi}g\right) + \frac{1}{2}, \qquad (2.12)$$

$$N > \frac{2}{\pi}g - \frac{1}{2\pi}\log\left(\frac{4}{\pi}g\right) - 1, \qquad (2.13)$$

while those of the second type must be evaluated numerically. In this case all the previously known limits can as well be computed analytically: BS: $N \leq g^2$; CC: $N \leq 4/\pi\, g$; M: $N \leq 2^{1/4}\, g$; C: $N \geq 2/\left(\pi\sqrt{e}\right)g - 1/2$; $C_0$: $N \geq g/\pi - 1/2$.

**Figure 2**. *Comparison between the exact number of bound states for the E potential (2.3) (ladder curve), the bounds of the first type (1.18) and (1.21), the C and $C_0$ lower bounds, the BS, CC and M upper bounds, and the bounds of the second type (1.26) and (1.29).*

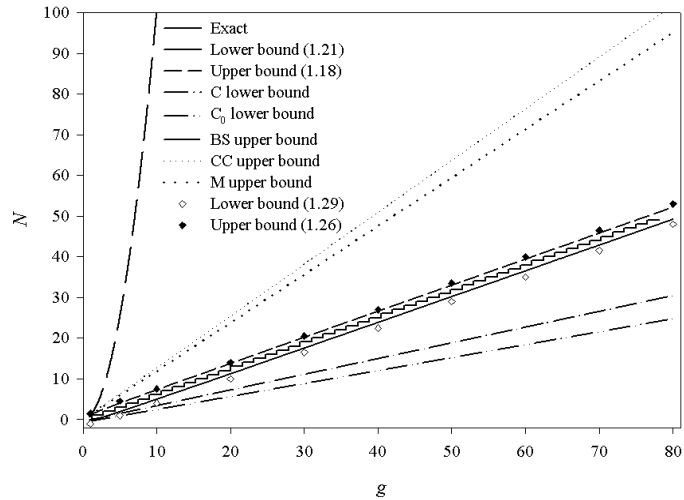

A comparison between these results is presented in Fig. 2. It is again clear that the new limits are remarkably effective.

The fourth test is performed with the (singular) H potential (2.4). In this case the exact number of bound states is given simply by the integer part of $g$:

$$N = \{\!\{g\}\!\}. \tag{2.14}$$

The new upper respectively lower limits of the first type applicable to singular potentials, (1.16) respectively (1.22), can in this case be computed analytically as well:

$$N \leq g - \frac{1}{\pi}\log\left(\tan\frac{\pi}{4g}\right) + \frac{1}{2}, \tag{2.15a}$$

$$N > g + \frac{1}{\pi}\log\left(\tan\frac{\pi}{4g}\right) - \frac{3}{2}, \tag{2.16a}$$

yielding asymptotically, for large $g$,

$$N \leq g - \frac{1}{\pi}\log\left(\frac{\pi}{4g}\right) + \frac{1}{2} - \frac{\pi}{48g^2} + O(g^{-4}), \tag{2.15b}$$

$$N > g + \frac{1}{\pi}\log\left(\frac{\pi}{4g}\right) - \frac{3}{2} + \frac{\pi}{48g^2} + O(g^{-4}). \tag{2.16b}$$

The new lower limit of the second type (1.29) must in this case be evaluated numerically, while all the previously known limits (relevant to the case of singular potentials) can be computed analytically: BS: $N \leq (\pi^2/6)g^2$; CC: $N \leq 2g$; C: $N \geq (2/\pi)\log(2)g - 1/2$.

**Figure 3**. *Comparison between the exact number of bound states for the H potential (2.4) (ladder curve), the bounds of the first type (1.16) and (1.22), the C lower bound, the BS and CC upper bounds, and the lower bound of the second type (1.29).*

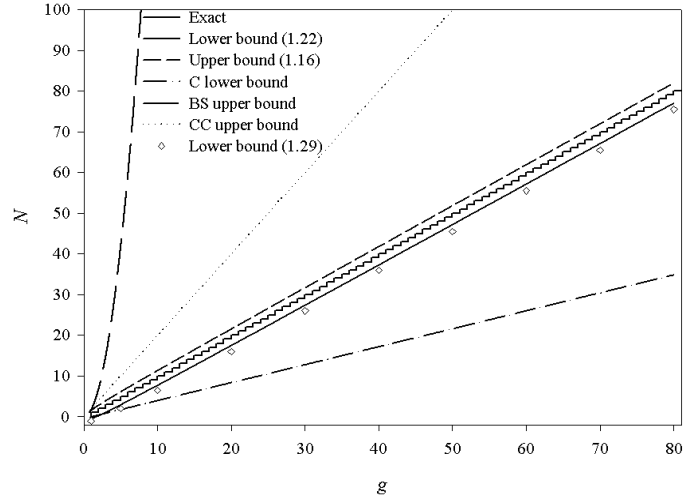

A comparison between these results is presented in Fig. 3. It is again clear that the new limits are remarkably effective. And it is again clear from a comparison of the asymptotic formulas (2.15b) and (2.16b) that the new upper and lower limits of the first type remain remarkably cogent even at large values of $g$: for instance, when the exact number of bound states is equal to $N = 5000$, these limits, (2.15a) and (2.16a), restrict $N$ to the relatively small interval $[4996, 5003]$. For comparison, the corresponding value of the BS upper limit exceeds $4 \times 10^7$, the CC upper limit is $10^4$, and the lower limit C only informs us that $N \geq 2207$; while the new lower limit of the second type, (1.29), informs us that $N \geq 4994$.

The fifth test is performed with the (singular) Y potential (2.5). In this case the exact number of bound states must be evaluated numerically: we employed two different methods of calculation, in order to check the numerical results [30,31] (note that these two methods possess a natural link [32]). The new upper and lower limits (1.16) respectively (1.22) can instead be computed analytically:

$$N \leq \sqrt{\frac{2}{\pi}} g + \frac{x^2 - y^2}{2\pi} + \frac{1}{2\pi} \log\left(\frac{x}{y}\right) + \frac{1}{2}, \tag{2.17a}$$

$$N > \sqrt{\frac{2}{\pi}} g - \frac{x^2 - y^2}{2\pi} - \frac{1}{2\pi} \log\left(\frac{x}{y}\right) - \frac{3}{2}, \tag{2.18a}$$

with

$$\text{erf}(y) = \alpha, \quad \text{erf}(x) = 1 - \alpha, \quad \alpha = (\pi/8)^{1/2} g^{-1}, \tag{2.19}$$

so that asymptotically (as $g \to \infty$, and keeping only the first correction term)

$$N \leq \sqrt{\frac{2}{\pi}} g + \frac{1}{\pi} \log(g), \tag{2.17b}$$

$$N > \sqrt{\frac{2}{\pi}} g - \frac{1}{\pi} \log(g). \tag{2.18b}$$

The new lower bound of the second type must also be evaluated numerically, while the previously known limits relevant to the singular case can all be evaluated (almost completely) analytically: BS: $N \leq g^2$; CC: $N \leq 2(2/\pi)^{1/2} g$; C: $N \geq (2/\pi) x^{1/2} \exp(-x/2) g - 1/2 = 0.531(2/\pi) g - 1/2$ (where $x$ is the root of $\exp(-x) = \int_x^\infty dy\, y^{-1} \exp(-y)$ ).

**Figure 4**. *Comparison between the exact number of bound states for the Y potential (2.4), the bounds of the first type (1.16) and (1.22), the C lower bound, the BS and CC upper bounds and the lower bound of the second type (1.29).*

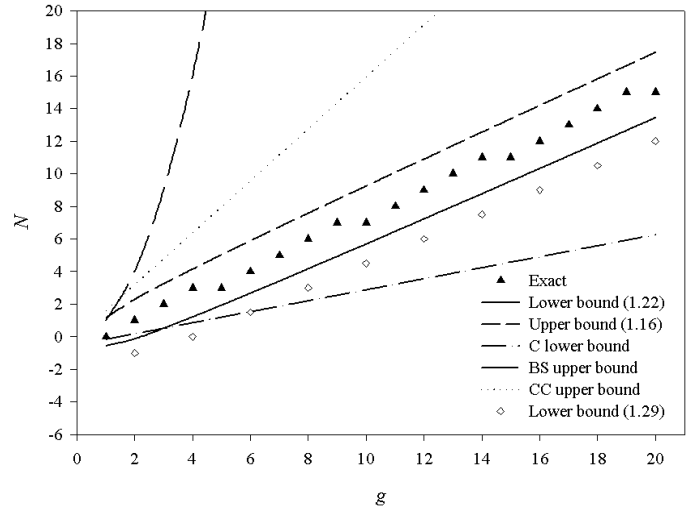

A comparison between these results is presented in Fig. 4. It is again clear that the new limits are remarkably effective. And it is again clear from a comparison of the asymptotic formulas (2.17b) and (2.18b) that the new upper and lower limits of the first type remain remarkably cogent even at large values of $g$: for instance, when the exact number of bound states is equal to $N = 50$, these limits, (2.17a) and (2.18a), restrict $N$ to the relatively small interval $[49,53]$. For comparison, the corresponding value of the BS upper limit exceeds 4000, while the CC upper limit and the C lower limit only informs us that $22 \leq N \leq 103$; as for the new lower limit, (1.29), it entails that $N \geq 48$.

Finally, the sixth test is performed with the (regular) STIS potential (2.6). As already mentioned, this test potential is particularly appealing because in this case *all* the relevant calculations can be performed analytically; moreover, in contrast to the five previous cases, this potential features two dimensionless parameters rather than only one. This potential possesses bound states only if the "coupling constant" $g$ exceeds one half, $g > 1/2$ (irrespective of the value of the other, *positive*, parameter it features, $\alpha > 0$), and the exact number $N$ of its bound states is then given again by (2.7), but now with

$$v = \frac{1}{2\pi}\left(\lambda \log(1+\alpha) + 2\arctan(\lambda)\right) \tag{2.20a}$$

with $\lambda = \sqrt{4g^2 - 1}$, entailing at large $g$

$$v = \frac{1}{\pi}\left(g - \frac{1}{8g}\right)\log(1+\alpha) - \frac{1}{2\pi g} + \frac{1}{2} + O(g^{-3}), \tag{2.20b}$$

and at large $\alpha$

$$v = \frac{\lambda}{2\pi}\left[\log(\alpha) + \frac{1}{\alpha}\right] + \frac{1}{\pi}\arctan(\lambda) + O(\alpha^{-2}). \tag{2.20c}$$

The new upper and lower limits of the first type (1.18) respectively (1.21) yield

$$\{\{v_{lo}\}\} \leq N \leq \{\{v_{up}\}\} \tag{2.21a}$$

with

$$v_{up} = \frac{1}{\pi}\left(g + \frac{1}{2}\right)\log(1+\alpha) - \frac{1}{4g} + \frac{1}{2}, \tag{2.21b}$$

respectively

$$v_{lo} = \frac{1}{\pi}\left(g - \frac{1}{2}\right)\log(1+\alpha) + \frac{1}{4g}, \tag{2.21c}$$

entailing at large $g$

$$\frac{v_{up} - v_{lo}}{v} = \frac{1}{g}\left[1 + \frac{\pi}{2\log(1+\alpha)}\right] - \frac{1}{g^2}\left[\frac{\pi}{\log(1+\alpha)} + \frac{\pi^2}{4\log^2(1+\alpha)}\right] + O(g^{-3}), \tag{2.21d}$$

and at large $\alpha$

$$\frac{v_{up} - v_{lo}}{v} = \frac{2}{\lambda}\left\{1 + \frac{\pi}{2\log(\alpha)}\left[1 - \frac{1}{g} - \frac{4}{\pi\lambda}\arctan(\lambda)\right]\right\} + O\left([\log(\alpha)]^{-2}\right). \tag{2.21e}$$

The new upper and lower limits of the second type (1.26) respectively (1.29) yield

$$\{\{v_-\}\} \leq N \leq \{\{v_+\}\} \tag{2.22a}$$

with

$$v_\pm = \frac{1}{2}\left\{\left\{\frac{2}{\pi}g_\pm \log(1+\alpha) - \frac{g_\pm}{g}\right\}\right\} + \frac{3 \pm 3}{4} \tag{2.22b}$$

where

$$g_\pm = \frac{\pm(\pi/2)}{\log[1 \pm \pi/(2g)]} \tag{2.22c}$$

so that, at large $g$,

$$g_{\pm} = g \pm \frac{\pi}{4} + O(g^{-1}) \qquad (2.22d)$$

entailing

$$v_{\pm} = \frac{1}{2}\left\{\left\{\left(\frac{2}{\pi}g \pm \frac{1}{2}\right)\log(1+\alpha)\right\}\right\} + \frac{1 \pm 3}{4} + O(g^{-1}) \qquad (2.22e)$$

hence

$$\frac{v_+ - v_-}{v} = O(g^{-1}). \qquad (2.22f)$$

Note that the formulas for the lower limit are only applicable if $g > \pi/2$ (see (2.22c)).

**Table 1**. *Comparison between the exact number of bound states $N$, the bounds of the first type (1.18) and (1.21), see (2.21), the limits of the second type (1.26) and (1.29), see (2.22), the BS, CC and M upper limits, and the C and $C_0$ lower limits, for the STIS potential (2.6) with a representative set of values of $\alpha$ and $g$.*

| $(\alpha, g)$ | $N$ | $\{\{v_{lo}\}\}$ | $\{\{v_{up}\}\}$ | $\{\{v_-\}\}$ | $\{\{v_+\}\}$ | BS | CC | M | C | $C_0$ |
|---|---|---|---|---|---|---|---|---|---|---|
| $(1,10)$ | 2 | 2 | 2 | 2 | 4 | 19 | 4 | 4 | 2 | 2 |
| $(1,10^2)$ | 22 | 21 | 22 | 22 | 24 | 1931 | 44 | 48 | 19 | 16 |
| $(1,10^3)$ | 221 | 220 | 221 | 220 | 222 | $>10^5$ | 441 | 488 | 186 | 159 |
| $(10^2,10)$ | 15 | 13 | 15 | 13 | 17 | 362 | 29 | 30 | 6 | 3 |
| $(10^2,10^2)$ | 147 | 146 | 148 | 146 | 150 | 36250 | 293 | 308 | 57 | 32 |
| $(10^4,10)$ | 29 | 27 | 31 | 27 | 33 | 821 | 58 | 99 | 6 | 3 |
| $(10^4,10^2)$ | 293 | 291 | 295 | 291 | 297 | 82105 | 586 | 999 | 63 | 32 |
| $(10^6,10)$ | 44 | 41 | 46 | 40 | 49 | 1281 | 87 | 316 | 6 | 3 |
| $(10^6,10^2)$ | 440 | 437 | 442 | 436 | 445 | $>10^5$ | 879 | 3162 | 64 | 32 |

The previously known upper and lower limits can also be evaluated in closed form for this potential:

$$\text{BS: } N \leq g^2\left[\log(1+\alpha) - \frac{\alpha}{1+\alpha}\right], \qquad (2.23)$$

$$\text{CC: } N \leq \frac{2}{\pi} g \log(1+\alpha), \qquad (2.24)$$

$$\text{M: } N \leq g\left[\left(\alpha - 2\log(1+\alpha) + \frac{\alpha}{1+\alpha}\right)\frac{\alpha}{1+\alpha}\right]^{1/4}, \qquad (2.25)$$

$$\text{C: } N \geq \frac{2}{\pi} g\left(1 - \frac{1}{\sqrt{1+\alpha}}\right) - \frac{1}{2}, \qquad (2.26)$$

$$\text{C}_0\text{: } N \geq \frac{1}{\pi} g \frac{\alpha}{1+\alpha} - \frac{1}{2}. \qquad (2.27)$$

The merits of the new limits are already apparent from these formulas. Representative examples are given in Table 1.

In conclusion it seems justified to conclude from these tests that the new limits presented in this paper are rather cogent and generally superior to those hitherto known. They are particularly effective for strong potentials possessing many bound states, thanks to their capability to generally reproduce the correct asymptotic (semiclassical) result (1.6) when the coupling constant diverges. Let us also emphasize that, from a computational point of view, the limits of the second type presented herein are particularly convenient, especially in the case of regular potentials.

**III. PROOFS**

In this Section we prove the new results reported in the introductory Section I. We assume throughout that the potential satisfies the conditions (1.1) as well as (1.3).

Let $u(r)$ be the zero-energy S-wave Schrödinger wave function, characterized by the second-order ordinary differential equation

$$u''(r) - V(r)u(r) = 0, \qquad (3.1a)$$

with boundary condition

$$u(0) = 0. \qquad (3.1b)$$

It is well known (see, for instance, [24]) that the number of zeros of the solution of (3.1a) with (3.1b) in the interval $0 < r < \infty$ coincides with the number $N$ of S-wave bound states supported by the potential $V(r)$ (we always exclude, for simplicity, the marginal case of a potential that features a "zero-energy bound state or resonance"). Let us indicate with $z_n$ the successive zeros of $u(r)$, and with $b_n$ the successive zeros of $u'(r)$ (namely, the locations of the successive extrema of the wave function $u(r)$),

$$u(z_n) = 0, \ u'(b_n) = 0. \qquad (3.2)$$

It is then clear that, since the potential $V(r)$ is nowhere *positive* (as implied by (1.1) with (1.3)),

$$V(r) = -|V(r)|, \qquad (3.3)$$

the zero-energy wave function $u(r)$ is an everywhere convex function of $r$, entailing the "interlacing" relations

$$0 = z_0 < b_1 < z_1 < b_2 < \cdots < z_{N-1} < b_N < z_N < \infty. \qquad (3.4)$$

Note that these formulas imply that $u'(r)$ does *not* vanish in the interval $z_N \leq r < \infty$, namely a $b_{N+1} < \infty$ does not exist (otherwise it would be inevitably followed by $z_{N+1} < \infty$, and this is excluded since $N$ is the number of zeros of $u(r)$).

Following [8, 24] we now introduce a function $\eta(r)$ defined via the relation

$$\tan[\eta(r)] = |V(r)|^{1/2} u(r)/u'(r), \qquad (3.5a)$$

with

$$\eta(0) = 0, \qquad (3.5b)$$

and the requirement that $\eta(r)$ be a continuous function of $r$ (to lift the $\mathrm{mod}(\pi)$ ambiguity entailed by the definition (3.5a)). It is then clear that the properties (3.4) together with the definition (3.5a) imply the relations

$$\eta(z_n) = n\pi, \ \eta(b_{n+1}) = (2n+1)\pi/2, \ n = 0,1,\ldots,N-1, \qquad (3.6a)$$

$$\eta(z_N) = \eta(\infty) = N\pi, \qquad (3.6b)$$

and that the value of $\eta(r)$ inside the intervals (3.4) lies between the values taken at the extremal points of these intervals, namely, for $z_n \leq r \leq b_{n+1}$ with $n = 0,\ldots,N-1$, $n\pi \leq \eta(r) \leq (2n+1)\pi/2$, and for $b_n \leq r \leq z_n$ with $n = 1,\ldots,N$, $(2n-1)\pi/2 \leq \eta(r) \leq n\pi$, except of course for the last interval, $z_N \leq r < \infty$, where $N\pi \leq \eta(r) < (2N+1)\pi/2$. Note that these results also imply that, for *all* values of $r$,

$$0 \leq \eta(r) < \left(N + \frac{1}{2}\right)\pi \qquad (3.6c)$$

(indeed the value at which the second inequality were violated would qualify as $b_{N+1}$, which, as already noted, would then inevitably be followed by $z_{N+1}$, violating the hypothesis that the number of zeros be $N$).

Moreover from (3.1a) we obtain via (3.5a) and (3.3) the nonlinear first-order differential equation

$$\eta'(r) = |V(r)|^{1/2} - \frac{V'(r)}{4|V(r)|} \sin[2\eta(r)], \tag{3.7}$$

which, together with the "initial condition" (3.5b), determines the function $\eta(r)$ and therefore, via (3.6b), the number $N$ of S-wave bound states. This equation will be our main tool to derive (upper and lower) limits on $N$.

It is indeed clear from (3.7) and (1.3) that

$$\eta'(r) \leq |V(r)|^{1/2} + \frac{V'(r)}{4|V(r)|}, \tag{3.8}$$

$$\eta'(r) \geq |V(r)|^{1/2} - \frac{V'(r)}{4|V(r)|}. \tag{3.9}$$

These inequalities, (3.8) respectively (3.9), together with (3.5b) and (3.6), will be our main tool to derive upper respectively lower limits on $N$. (Note that more stringent conditions might be written by considering separately all the intervals of type $z_n \leq r \leq b_{n+1}$ where $\sin[2\eta(r)]$ is clearly *nonnegative*, see (3.4) and (3.6a), respectively all the intervals of type $b_n \leq r \leq z_n$ where $\sin[2\eta(r)]$ is clearly *nonpositive*, see (3.4) and (3.6a); but it does not appear that such a distinction might be maintained to the end without having to renounce the goal to obtain reasonably neat final formulas for the limits; we will however take advantage of this improvement for certain intervals, see below).

Let us now focus firstly on the derivation of the upper limit (1.16). To this end we integrate (3.8) from $b_1$ to $z_{N-1}$, and via (3.6a) and (3.3) we get

$$\left(N - \frac{3}{2}\right)\pi \leq \int_{b_1}^{z_{N-1}} dr\, |V(r)|^{1/2} + \frac{1}{4}\log\left|\frac{V(b_1)}{V(z_{N-1})}\right|. \tag{3.10}$$

On the other hand we know, as already noted above, that in the intervals $0 \leq r \leq b_1$ and $z_{N-1} \leq r \leq b_N$ (where $\sin[2\eta(r)]$ is *nonnegative*, see (3.4) and (3.6a)) (3.8) can be replaced by the more stringent inequality (see (3.7))

$$\eta'(r) \leq |V(r)|^{1/2}, \tag{3.11a}$$

and the integration of this inequality over these intervals yields (via (3.6a))

$$\frac{\pi}{2} \leq \int_0^{b_1} dr\, |V(r)|^{1/2}. \tag{3.11b}$$

$$\frac{\pi}{2} \leq \int_{z_{N-1}}^{b_N} dr\, |V(r)|^{1/2}. \tag{3.11c}$$

Hence by summing (3.10), (311b) and (3.11c) (and dividing by $\pi$) we get

$$N - \frac{1}{2} \leq \frac{1}{\pi} \int_0^{b_N} dr\, |V(r)|^{1/2} + \frac{1}{4\pi} \log\left|\frac{V(b_1)}{V(z_{N-1})}\right|, \tag{3.12}$$

and therefore *a fortiori* (thanks to the monotonicity of $V(r)$, see (1.3))

$$N \leq \frac{1}{\pi} \int_0^{\infty} dr\, |V(r)|^{1/2} + \frac{1}{4\pi} \log\left|\frac{V(p)}{V(q)}\right| + \frac{1}{2}, \tag{3.13a}$$

provided

$$p \leq b_1, \tag{3.13b}$$

$$q \geq z_{N-1}. \tag{3.13c}$$

To complete the proof of the first upper limit reported in Section I, see (1.16), we must show that the radii $p$ respectively $q$ defined by (1.16b) respectively (1.16c) satisfy (3.13b) respectively (3.13c). For $p$ this is immediately implied by a comparison of (1.16b) and (3.11b); and likewise, indeed *a fortiori*, this is as well implied for $q$ by a comparison of (1.16c) and (3.11c).

Let us now proceed and prove the first lower limit of Section I. We treat firstly the case in which the potential is finite at the origin, see (1.19). To this end we integrate (3.9) from $0$ to an arbitrary (of course *positive*) radius $s$, getting thereby the inequality

$$\eta(s) \geq \int_0^s dr |V(r)|^{1/2} - \frac{1}{4}\log\left|\frac{V(0)}{V(s)}\right|, \tag{3.14}$$

namely *a fortiori*, via (3.6c),

$$\left(N + \frac{1}{2}\right)\pi > \int_0^s dr |V(r)|^{1/2} - \frac{1}{4}\log\left|\frac{V(0)}{V(s)}\right|, \tag{3.15}$$

which clearly immediately implies (1.19).

If the potential diverges at the origin, to get the lower bound (1.22) we integrate (3.9) from $p$ to $q$, and we then get via (1.16b,c)

$$\eta(q) - \eta(p) \geq \int_p^q dr |V(r)|^{1/2} - \frac{1}{4}\log\left|\frac{V(p)}{V(q)}\right| = \int_0^\infty dr |V(r)|^{1/2} - \pi - \frac{1}{4}\log\left|\frac{V(p)}{V(q)}\right|, \tag{3.16}$$

and via (3.6c) this clearly yields (1.22).

A generally more stringent but less explicit bound obtains by integrating (3.9) from $b_1$ to $s$, getting thereby (see (3.6a))

$$\eta(s) - \frac{\pi}{2} \geq \int_{b_1}^s dr |V(r)|^{1/2} - \frac{1}{4}\log\left|\frac{V(b_1)}{V(s)}\right|, \tag{3.17a}$$

hence *a fortiori*, via (3.6c),

$$N\pi > \int_{b_1}^s dr |V(r)|^{1/2} - \frac{1}{4}\log\left|\frac{V(b_1)}{V(s)}\right|, \tag{3.17b}$$

hence *a fortiori* (see (3.13b) and (1.3))

$$N\pi > \int_{b_1}^s dr |V(r)|^{1/2} - \frac{1}{4}\log\left|\frac{V(p)}{V(s)}\right|, \tag{3.18}$$

hence finally

$$N\pi > \int_t^s dr |V(r)|^{1/2} - \frac{1}{4}\log\left|\frac{V(p)}{V(s)}\right|, \tag{3.19a}$$

provided there holds the inequality

$$t \geq b_1. \tag{3.19b}$$

This condition is clearly equivalent to the requirement that the potential $V(r)$ amputated of its part extending beyond $t$ possess at least one bound state (since when $V(r)$ vanishes, $u(r)$ is linear, $u(r) = \alpha r + \beta$, see (3.1a), hence the condition (3.19b) with $V(r)$ vanishing beyond $t$ guarantees the existence of $z_1 < \infty$). It is therefore sufficient, to make sure that (3.19b) hold, that this amputated potential, $V(r)\theta(t-r)$ (where $\theta(x)$ is the step function, $\theta(x) = 1$ if $x \geq 0$, $\theta(x) = 0$ if $x < 0$) satisfy one of the *sufficient* conditions for the existence of at least one bound state reported in Section I, see (1.11-13). Here for simplicity we restrict attention to the *sufficient* condition (1.12), and we thereby conclude that a formula adequate to guarantee that the inequality (3.19b) be satisfied is validity, for some *positive* value of $a$, of either one of the following two inequalities, see (1.12) (below we write $\geq$ in place of $>$, since $t$ might coincide with $b_1$, see (3.19b), which would correspond to an amputated potential possessing only a zero-energy bound state or resonance):

$$a^{-1} \int_0^a dr\, r^2 |V(r)| + a \int_a^t dr\, |V(r)| \geq 1 \quad \text{with} \quad a \leq t, \qquad (3.20a)$$

$$a^{-1} \int_0^a dr\, r^2 |V(r)| \geq 1 \quad \text{with} \quad a \geq t. \qquad (3.20b)$$

And clearly the choice $a = t$ leads to (1.23b), thereby completing the proof of the first lower limit to $N$ for potentials singular at the origin as reported in Section I, see (1.23).

Let us now proceed and prove the second type of limits to $N$. For simplicity, in the case of the upper bound we restrict attention to the case of potentials which are finite at the origin, and of course we always assume the potential to satisfy the monotonicity condition (1.3).

First of all we introduce the potential amputated of its part beyond $q$,

$$\overline{V}(r) = V(r) \quad \text{for} \quad 0 \leq r < q, \qquad (3.21a)$$

$$\overline{V}(r) = 0 \quad \text{for} \quad r \geq q. \qquad (3.21b)$$

Here $q$ is defined by (1.16c), hence it satisfies the condition (3.13c); therefore, if we indicate with $\overline{N}$ the number of bound states possessed by the potential $\overline{V}(r)$, either $\overline{N} = N - 1$ (if $z_{N-1} \leq q < b_N$; indeed the zero-energy wave function $\overline{u}(r)$ corresponding to the potential $\overline{V}(r)$ is linear for $r > q$, see (3.21b), hence it has one less zero than the zero-energy wave function $u(r)$ corresponding to the potential $V(r)$ if the cutoff point $q$ comes before the point, $b_N$, at which $u(r)$ bends over for the last time, namely where it has its last extremum) or $\overline{N} = N$ (if $q \geq b_N$; we include in the count of the number $\overline{N}$ of bound states of $\overline{V}(r)$ also a zero-energy one, should it happen that there be one, namely that $q = b_N$). So, in any case

$$\overline{N} \leq N \leq \overline{N} + 1. \qquad (3.22)$$

Our strategy is now to introduce two monotonically increasing ladder-type potentials, $V^{(+)}(r)$ respectively $V^{(-)}(r)$, both vanishing beyond $q$ just as $\overline{V}(r)$ does (see (3.21b)), which minorize respectively majorize $\overline{V}(r)$,

$$V^{(+)}(r) \leq \overline{V}(r) \leq V^{(-)}(r), \qquad (3.23)$$

so that the number of bound states, $N^{(+)}$ respectively $N^{(-)}$, possessed by them majorize respectively minorize $\overline{N}$, yielding, via (3.22),

$$N^{(-)} \leq N \leq N^{(+)} + 1. \qquad (3.24)$$

And these potentials, $V^{(+)}(r)$ respectively $V^{(-)}(r)$, shall now be manufactured so that one can easily compute the numbers of bound states they possess.

Indeed the potential $V^{(+)}(r)$ is now defined by the rule

$$V^{(+)}(r) = V\left(r_j^{(+)}\right) \quad \text{for} \quad r_j^{(+)} \leq r < r_{j+1}^{(+)}, \quad j = 0, 1, \ldots, J^{(+)} - 1, \qquad (3.25a)$$

$$V^{(+)}(r) = V\left(r_{J^{(+)}}^{(+)}\right) \quad \text{for} \quad r_{J^{(+)}}^{(+)} \leq r < q, \qquad (3.25b)$$

$$V^{(+)}(r) = 0 \quad \text{for} \quad r \geq q, \qquad (3.25c)$$

with the increasing radii $r_j^{(+)}$ defined by the recurrence relation (1.24), and the *positive* integer $J^{(+)}$ defined by the condition that the radius $r_{J^{(+)}+1}^{(+)}$ yielded by this recursion (be the first one to) exceed or equal $q$, see (1.25). It is plain that this potential minorizes, see (3.23), the truncated potential $\overline{V}(r)$ for all values of $r$ (if in doubt, draw a graph), and it is moreover easy to compute the number $N^{(+)}$ of bound states it possesses, since for this potential

$$\eta^{(+)}\left(r_j^{(+)}\right) = j\pi/2, \quad j = 0,1,\ldots,J^{(+)} + 1. \tag{3.26}$$

This result is implied by the differential equation satisfied by $\eta^{(+)}(r)$, which reads simply

$$\eta'^{(+)}(r) = \left|V^{(+)}(r)\right|^{1/2}, \tag{3.27a}$$

namely (see (3.25a))

$$\eta'^{(+)}(r) = \left|V^{(+)}\left(r_j^{(+)}\right)\right|^{1/2} \quad \text{for} \quad r_j^{(+)} \le r < r_{j+1}^{(+)}, \quad j = 0,1,\ldots,J^{(+)}, \tag{3.27b}$$

since the second term in the right-hand side of (3.7) vanishes for $r_j < r < r_{j+1}$ because $V^{(+)}(r) = V^{(+)}\left(r_j^{(+)}\right)$ is constant there hence its derivative vanishes, and at $r = r_j$ because $\sin\left[2\eta^{(+)}\left(r_j^{(+)}\right)\right]$ vanishes due to (3.26) and therefore kills the contribution that would otherwise come from the delta function produced by the derivative of the discontinuity of the potential occurring there. And the consistency of (3.26) with (3.27) is of course guaranteed by (3.25b) and (1.24).

We now note that, for this potential $V^{(+)}(r)$, (3.26) implies

$$r_j^{(+)} = z_{j/2}^{(+)} \quad \text{if } j \text{ is } even, \quad j = 0,2,\ldots,J^{(+)} - 1 \text{ or } J^{(+)}, \tag{3.28a}$$

$$r_j^{(+)} = b_{(j+1)/2}^{(+)} \quad \text{if } j \text{ is } odd, \quad j = 1,3,\ldots,J^{(+)} - 1 \text{ or } J^{(+)}, \tag{3.28b}$$

where the radii $z_j^{(+)}$ respectively $b_j^{(+)}$ are of course the successive zeros respectively the extrema of the zero-energy wave function $u^{(+)}(r)$ corresponding to the potential $V^{(+)}(r)$ (see (3.4)). Moreover, for a potential amputated of its part beyond $q$ (as is the case of $V^{(+)}(r)$), the number $N^{(+)}$ of bound states is characterized by the condition $b_{N^{(+)}}^{(+)} \le q$ (since the zero-energy wave function is a straight line for $r > q$, see (3.1a) and (3.25c)). Hence after considering the two possible parities, even or odd, of $J^{(+)}$, we conclude that, in both cases,

$$N^{(+)} = \left\{\!\left\{\left(J^{(+)} + 1\right)\!/2\right\}\!\right\}, \tag{3.29}$$

and via (3.24) this completes our proof of the upper limit (1.26).

To prove the lower limit (1.29) we introduce the following ladder-type potential:

$$V^{(-)}(r) = V\!\left(r_{J^{(-)}}^{(-)}\right) \quad \text{for} \quad 0 \le r \le r_{J^{(-)}}^{(-)}, \tag{3.30a}$$

$$V^{(-)}(r) = V\!\left(r_{j-1}^{(-)}\right) \quad \text{for} \quad r_j^{(-)} < r \le r_{j-1}^{(-)}, \quad j = J^{(-)}, J^{(-)} - 1,\ldots,2,1, \tag{3.30b}$$

$$V^{(-)}(r) = 0 \quad \text{for} \quad q = r_0^{(-)} < r < \infty, \tag{3.30c}$$

with the sequence of *decreasing* radii $r_j^{(-)}$ defined by the recursion relation (1.27). It is plain that this potential majorizes, see (3.23), the truncated potential $\overline{V}(r)$ for all values of $r$ (if in doubt, draw a graph); hence if $N^{(-)}$ is the number of S-wave bound states possessed by this potential, the (first part of the) inequality (3.24) holds. As we know, since the potential $V^{(-)}(r)$ vanishes identically beyond $q$ ($b_{N^{(-)}}^{(-)} \le q$), see (3.30c), this number $N^{(-)}$ is given by

$$N^{(-)} = \left\{\!\left\{\eta^{(-)}(q)/\pi\right\}\!\right\}. \tag{3.31}$$

Here $\eta^{(-)}(r)$ is of course the solution of the differential equation (3.7) for the potential $V^{(-)}(r)$, namely

$$\eta'^{(-)}(r) = \left|V^{(-)}(r)\right|^{1/2} - \frac{V'^{(-)}(r)}{4\left|V^{(-)}(r)\right|}\sin\left[2\eta^{(-)}(r)\right], \tag{3.32}$$

with the initial condition

$$\eta^{(-)}(0) = 0. \tag{3.33}$$

Since the ladder-type potential $V^{(-)}(r)$ presents some discontinuities, see (3.30), the integration of (3.32) from the initial condition (3.33) onward shall encounter some delta functions, but these integrable singularities of the right-hand side of (3.32) do not destroy the properties of existence, uniqueness and continuity of the solution $\eta^{(-)}(r)$ of (3.32) with (3.33).

Let now $\tilde{\eta}(r)$ be another solution of the same differential equation (3.32), characterized by the initial condition

$$\tilde{\eta}(0) = -r_{J^{(-)}}^{(-)}\left|V\left(r_{J^{(-)}}^{(-)}\right)\right|^{1/2}. \tag{3.34}$$

Since clearly (see (3.33) and (1.28))

$$\eta^{(-)}(0) > \tilde{\eta}(0) \tag{3.35a}$$

and the two functions $\eta^{(-)}(r)$ and $\tilde{\eta}(r)$ satisfy the same differential equation, there follows that, for every finite value of $r$ an analogous inequality holds (indeed, the graph of the continuous function $\tilde{\eta}(r)$ as function of $r$ can never overtake the graph of the continuous function $\eta^{(-)}(r)$ as function of $r$, since at the point of crossing their slopes must coincide because $\tilde{\eta}(r)$ satisfies the same differential equation as $\eta^{(-)}(r)$, see (3.32), hence no crossing can occur):

$$\eta^{(-)}(r) > \tilde{\eta}(r). \tag{3.35b}$$

Hence as well

$$\eta^{(-)}(q) > \tilde{\eta}(q), \tag{3.35c}$$

entailing *a fortiori*, via (3.31),

$$N^{(-)} \geq \{\{\tilde{\eta}(q)/\pi\}\}. \tag{3.36}$$

(Note that, though a strict inequality sign appears in (3.35c), one must allow for the possibility of equality in this formula, (3.36), because two *different* numbers may have the *same* integer part).

But the initial condition (3.34), and the recursion relation (1.27) defining the radii $r_j^{(-)}$, have been adjusted, as it can be easily verified in analogy to the argument used above, so that $\tilde{\eta}\left(r_{J^{(-)}}^{(-)}\right) = 0$, $\tilde{\eta}\left(r_{J^{(-)}-1}^{(-)}\right) = \pi/2$, $\tilde{\eta}\left(r_{J^{(-)}-2}^{(-)}\right) = \pi$, and so on, entailing (see (1.27))

$$\tilde{\eta}\left(r_0^{(-)}\right) = \tilde{\eta}(q) = J^{(-)}\pi/2. \tag{3.37}$$

Via (3.31) and (3.24) this implies the lower limit (1.29), which is thereby proven.

## IV. THE KLEIN-GORDON CASE

In the context of first-quantized mechanics with relativistic kinematics, a zero-spin particle of (*positive*) mass $m$ moving in an external potential $W(\vec{r})$ which is the fourth-component of a relativistic 4-vector is described (in self-evident notation, and with an appropriate choice of units) by the following Klein-Gordon equation:

$$\left(\vec{P}^2 + m^2\right)\psi(\vec{r}) = \left[E - W(\vec{r})\right]^2 \psi(\vec{r}). \tag{4.1}$$

In the spherically-symmetrical case, $W(\vec{r}) = W(r)$, the zero-kinetic-energy (namely, $E = m$) S-wave radial equation coincides with the corresponding equation for the Schrödinger case, (3.1), with the following definition of $V(r)$ in terms of $W(r)$:

$$V(r) = 2mW(r) - W^2(r). \tag{4.2}$$

Note that, if the potential $W(r)$ is monotonically nondecreasing and vanishes at infinity (and is therefore *nonpositive*, $W(r) = -|W(r)|$), the same property, see (1.3), holds as well for the potential $V(r)$. And the following conditions on the behavior of $W(r)$ at the origin and at infinity are clearly sufficient to guarantee the validity of (1.1):

$$\lim_{r \to 0} \left[ r^{1-\varepsilon} W(r) \right] = 0, \tag{4.3a}$$

$$\lim_{r \to \infty} \left[ r^{2+\varepsilon} W(r) \right] = 0. \tag{4.3b}$$

All the results reported above in the Schrödinger context can therefore be immediately taken over to the Klein-Gordon case. Note however that, as a consequence of the relation (4.2), if one introduces a "coupling constant" $g$ as a measure of the strength of the potential by setting $W(r) = g^2 w(r)$, then one sees that in the Klein-Gordon case as $g$ diverges the number of S-wave bound states grows proportionally to $g^2$ (rather than proportionally to $g$ as is the case in the Schrödinger context, see (1.6)).

## ACKNOWLEDGMENTS

One of us (FB) would like to thank Professor F. Michel for its interest, Dr. C. Semay for checking some numerical results, and FNRS for financial support.

---